\documentclass[10pt,twocolumn,letterpaper]{article}

\usepackage[pagenumbers]{cvpr} 

\definecolor{cvprblue}{rgb}{0.21,0.49,0.74}
\usepackage[pagebackref,breaklinks,colorlinks,citecolor=cvprblue]{hyperref}

\usepackage{multirow}

\def\paperID{13281} 
\def\confName{CVPR}
\def\confYear{2025}

\title{From Faces to Voices: Learning Hierarchical Representations \\for High-quality Video-to-Speech}

\author{Ji-Hoon Kim \quad Jeongsoo Choi \quad Jaehun Kim \quad Chaeyoung Jung \quad  Joon Son Chung \\
Korea Advanced Institute of Science and Technology\\
{\tt\small \{jihoon, joon\}@mm.kaist.ac.kr}
}

\begin{document}
\maketitle
\begin{abstract}
The objective of this study is to generate high-quality speech from silent talking face videos, a task also known as video-to-speech synthesis.
A significant challenge in video-to-speech synthesis lies in the substantial modality gap between silent video and multi-faceted speech. 
In this paper, we propose a novel video-to-speech system that effectively bridges this modality gap, significantly enhancing the quality of synthesized speech.
This is achieved by learning of hierarchical representations from video to speech.
Specifically, we gradually transform silent video into acoustic feature spaces through three sequential stages -- content, timbre, and prosody modeling.
In each stage, we align visual factors -- lip movements, face identity, and facial expressions -- with corresponding acoustic counterparts to ensure the seamless transformation.
Additionally, to generate realistic and coherent speech from the visual representations, we employ a flow matching model that estimates direct trajectories from a simple prior distribution to the target speech distribution.
Extensive experiments demonstrate that our method achieves exceptional generation quality comparable to real utterances, outperforming existing methods by a significant margin.
\end{abstract}

\vspace{-1mm}
\section{Introduction}
\label{sec:intro}
Video-to-Speech (VTS) systems have recently attracted significant attention for their capability to convert silent videos of talking faces into human speech.
These systems have a broad spectrum of applications, such as re-dubbing silent archival films, providing assistive technologies for individuals with speech disabilities, and enabling natural communications in loud settings~\cite{choi2023diffv2s,kim2024let,yeminilipvoicer}.
Recent advancements in deep learning have propelled this field forward by utilizing the natural alignment of video and speech as a mode of training supervision, eliminating the need of additional annotations such as text transcriptions.

The ultimate goal of VTS systems is to synthesize realistic human speech.
A key challenge in building high-quality VTS systems lies in the significant information gap between silent video and spoken audio.
Specifically, silent video primarily contains visual features such as lip motion and facial expressions, while spoken audio includes acoustic characteristics such as tone and pronunciation.
As a result, the VTS systems require capturing the complex relations between the two modalities, making it difficult to establish an accurate mapping from visual to acoustic spaces.

\begin{figure}[t]
    \centering
    \vspace{5mm}
    \includegraphics[width=0.99\columnwidth]{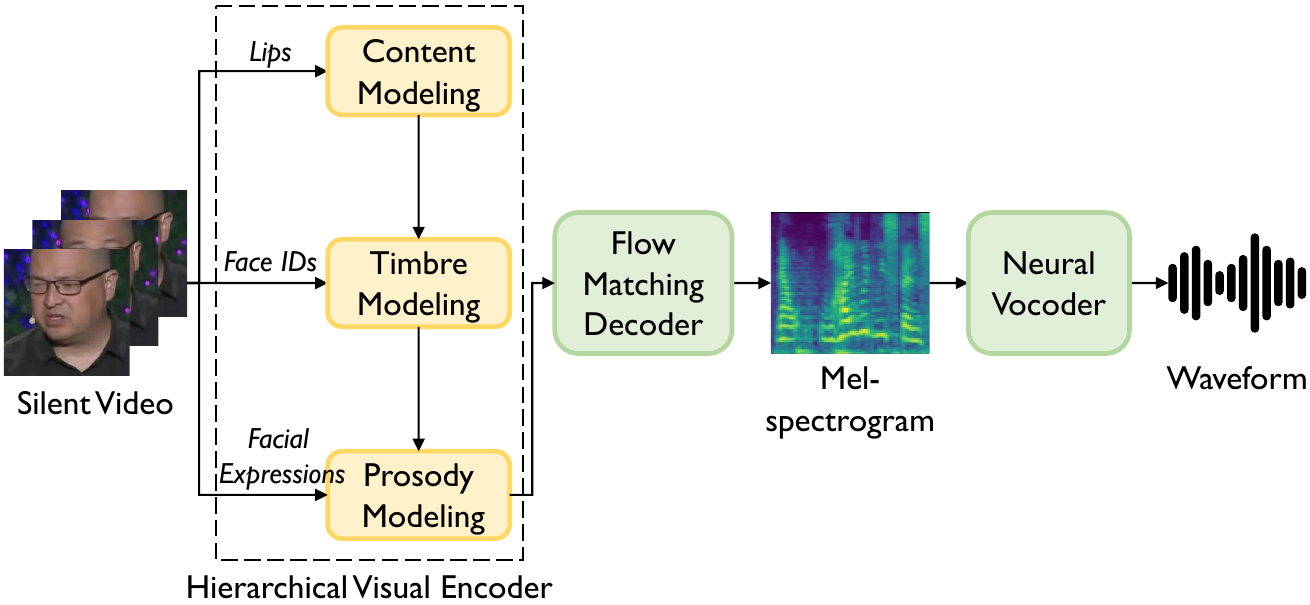}
    \caption{An overview of the proposed system. 
    Our method learns hierarchical representations from video to speech, focusing on three key factors: lips, face IDs, and facial expressions.
    The visual encoding is converted into the corresponding speech through an effective flow matching decoder and neural vocoder.
    }    
    \vspace{-5mm}
    \label{fig:teaser}
\end{figure}

Numerous works have been made to enhance the quality of VTS system while addressing the information gap between the two modalities.
To mitigate the complexity posed by the inherent variability of speech, some approaches incorporate self-supervised speech units~\cite{choi2023diffv2s,hsu2023revise} or a dedicated lip-reading network~\cite{yeminilipvoicer}.
Other studies have focused on clarifying multiple speaker characteristics by utilizing speaker embeddings extracted from either the reference audio~\cite{mira2022svts,prajwal2020learning} or the input video ~\cite{choi2023diffv2s,yeminilipvoicer}.
Meanwhile, several approaches adopt advanced modeling techniques, such as diffusion models~\cite{choi2023diffv2s,yeminilipvoicer,zheng2024speech}, to capture the intricate relationships between visual and acoustic modalities.
Despite these efforts, current VTS systems still struggle to close the modality gap, leaving their generation quality significantly behind that of real human utterances.
This underscores the need for a more effective training pipeline and refined network architecture for high-quality VTS systems.

In this paper, we propose a novel video-to-speech approach that effectively bridges the gap between what the eyes see and what the ears can hear.
To achieve this, we design a hierarchical visual encoder that refines hierarchical representations from video to speech. 
Specifically, we divide VTS mapping into three stages--content, timbre, and prosody--and incrementally transform visual input into acoustic feature spaces. 
Since content exhibits less variation and directly influences both timbre and prosody~\cite{zhang2023learning,ren2022prosospeech}, we prioritize content modeling as the first stage. 
Timbre modeling is placed before prosody modeling, as timbre tends to be more stable than prosody~\cite{jiang2024towards}, and distinguishing timbre helps reduce ambiguity in prosody modeling~\cite{ren2022prosospeech}.

In addition, to facilitate seamless transformation at each stage, we align multiple visual cues with their acoustic counterparts (Figure~\ref{fig:teaser}).
For content modeling, we leverage lip motions, inspired by the strong correlation between lip movements and speech content~\cite{hsu2022u,shi2022learning}. 
For timbre modeling, we incorporate face identity as a conditional input, drawing from the cross-modal biometric correlation between facial appearance and timbre~\cite{nagrani2018seeing,gao2021visualvoice,ning2021disentangled}. 
Lastly, for prosody modeling, we utilize facial expressions, which convey emotions and subtle nuances, naturally aligning with pitch and energy variations in speech~\cite{toisoul2021estimation,cong2023learning}.

Based on the encoded visual representations which are adapted to the acoustic feature space, we aim to generate realistic mel-spectrograms.
To this end, we employ a flow matching generative decoder that offers a streamlined and accurate generation process, achieving high-fidelity speech synthesis in fewer sampling steps~\cite{lipman2022flow,mehta2024matcha}.
We conduct extensive experiments on two datasets collected from in-the-wild settings~\cite{son2017lip,afouras2018lrs3}.
The experimental results demonstrate that our method achieves exceptional audio quality, making a Mean Opinion Score (MOS) gap of only \textit{0.05} in naturalness, compared to the real human utterances.
The synthesized audio samples can be found on our demo page\footnote{\url{https://mm.kaist.ac.kr/projects/faces2voices}}.

\begin{figure*}[ht]
    \centering
    \includegraphics[width=.97\textwidth]{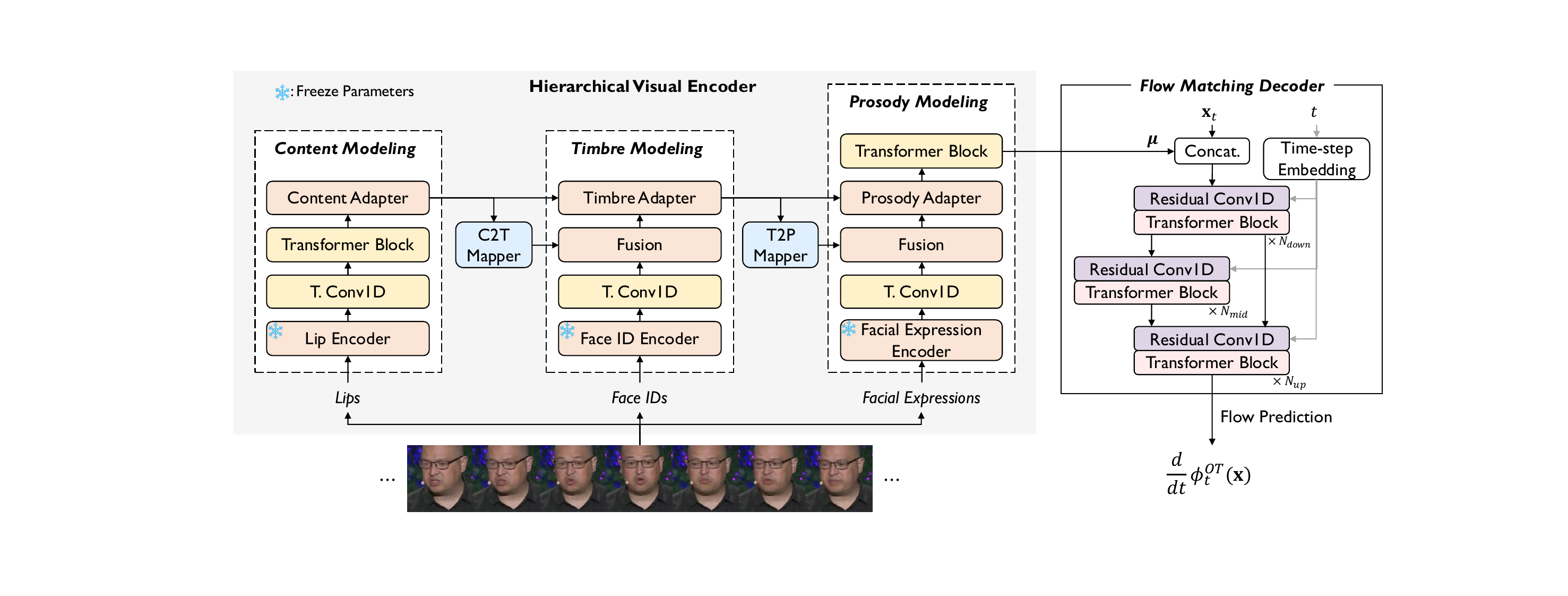}
    \caption{
    The detailed architecture of the our framework.
    Our approach gradually closes the substantial modality gap between video and speech, while aligning key visual cues--lip movements, face identity, and facial expressions--with their corresponding speech attributes--content, timbre, and prosody.
    The flow matching decoder effectively estimates mel-spectrogram distribution, conditioned on the visual encoding $\boldsymbol{\mu}$. 
    ${\bf x}_t$ represents an intermediate state of mel-spectrogram at time-step $t$, and $\phi_t^{OT}$ denotes the corresponding flow.
    }
    \label{fig:architecture}
\end{figure*}

\section{Related Works}
\subsection{Video-to-Speech}
VTS systems have experienced significant advancements, transitioning from rule-based approaches~\cite{le2015reconstructing,le2017generating} to contemporary end-to-end methods~\cite{yadav2021speech,wang2022fastlts}.
Early deep learning approaches predominantly employed convolution neural networks, demonstrating its effectiveness in VTS systems~\cite{ephrat2017vid2speech,kumar2019lipper}.
More recent studies have improved generation quality by incorporating advanced modeling techniques such as generative adversarial networks~\cite{kim2021lip,mira2022end}, normalizing flows~\cite{he2022flow,kim2024let}, and diffusion networks~\cite{choi2023diffv2s,yeminilipvoicer,zheng2024speech}.

Meanwhile, there have been efforts to utilize auxiliary information to mitigate the difference between video and speech data distributions.
To complement the lack of supervision from speech data itself, Kim et al.~\cite{kim2023lip} utilize text transcriptions as an auxiliary target.
More recent works~\cite{hsu2023revise,choi2023intelligible,kim2024let,lei2024uni} have adopted quantized self-supervised speech representations, eliminating the need of text transcriptions.
In order to capture multiple speaker characteristics, many works incorporate speaker embeddings derived from the reference audio~\cite{prajwal2020learning,mira2022svts,hegde2022lip,choi2023intelligible}. 
However, since obtaining reference audio is not always feasible during inference process, DiffV2S~\cite{choi2023diffv2s} introduces video-driven speaker embeddings focusing on lip frames, whereas LipVoicer~\cite{yeminilipvoicer} estimates speaker information through a single portrait image. 
In contrast to previous works, we focus directly on bridging the modality gap between video and speech, while associating multiple visual cues with their corresponding acoustic counterparts.

\subsection{Hierarchical Speech Generation}
Due to the inherent complexity of speech, various studies have explored hierarchical generative approaches for high-quality speech synthesis.
In the context of text-to-speech synthesis, Hsu et al.~\cite{hsuhierarchical} develop a system that uses two-tiered latent variable modeling based on a conditional variational autoencoder.
The first level captures coarse acoustic information, while the second level deals with specific attribute configurations.
Both PVAE-TTS~\cite{lee2022pvae} and Grad-StyleSpeech~\cite{kang2023grad} employ hierarchical structures in adaptive text-to-speech systems.
To address the challenges of mimicking new speaking styles, these systems improve their adaptation capabilities through a progressive variational autoencoder and a hierarchical encoder, respectively.
Similarly, HierVST~\cite{lee2023hiervst} adopts a hierarchical structure in their voice style transfer system. 
To effectively handle speaker styles not encountered during training, HierVST first generates linguistic information and then integrates it with residual acoustic information through hierarchical variational inference.
In our work, we explore hierarchical representations from silent video to human speech, and propose a high-quality VTS system that generates natural speech through these hierarchical representations.

\subsection{Flow Matching}
Flow matching~\cite{lipman2022flow} has recently gained increasing attention due to its capability to generate realistic data samples with straight trajectories, addressing the inherent slow sampling issues in diffusion-based models~\cite{ho2020denoising}.
The effectiveness of flow matching has been demonstrated across various research fields, including vision~\cite{fischer2023boosting,hu2024latent} and audio~\cite{liu2023generative,prajwal2024musicflow} domains.
In vision domain, Fischer et al.~\cite{fischer2023boosting} adopt a flow matching model between a frozen diffusion model and a convolution block, which enables effective image synthesis.
Similarly, Hu et al.~\cite{hu2024latent} utilize flow matching in their image editing pipeline, benefiting from its streamlined and efficient inference process.
In the realm of audio generation, 
SpeechFlow~\cite{liu2023generative} applies flow matching to build a robust foundation model for speech, showcasing powerful performance across diverse downstream tasks including speech separation and enhancement.
MusicFlow~\cite{prajwal2024musicflow} introduces a text-guided music generation method based on two flow matching networks to capture the conditional distribution of semantic and acoustic features.
Building on these advancements, our VTS system employs flow matching to bridge the visual-to-audio modality gap, resulting in the natural and intelligible generation of speech from silent video.

\section{Method}

\subsection{Overall Architecture}
As illustrated in Figure~\ref{fig:architecture}, our framework mainly consists of a hierarchical visual encoder and a flow matching decoder.
The hierarchical visual encoder gradually refines video representations, aligning visual cues--lip movements, face identity, and facial expressions--with their corresponding acoustic counterparts--content, timbre, and prosody.
This ensures a seamless transformation from visual to acoustic modalities, enhancing the naturalness and clarity of the synthesized speech.
The resulting visual encoding $\boldsymbol{\mu}$ are fed into a flow matching decoder, and then flow matching decoder generates a high-quality mel-spectrogram, which is subsequently converted into audible waveform by a pre-trained neural vocoder~\cite{kong2020hifi}. 

\subsection{Hierarchical Visual Encoder}
To effectively close the large gap between video and speech, we propose a hierarchical visual encoder that gradually transforms input videos into acoustic feature spaces, starting from a fundamental attribute and advancing to complex ones.
The visual encoder sequentially models content, timbre, and prosody, with mappers that enable interaction across these distinct modeling processes.
Each mapper is based on Transformer layers which facilitate better understanding of underlying sequences~\cite{gulati2020conformer}.

Inspired by the cross-modal correlations between face and speech, our visual encoder aligns lip motions, facial identity, and expressions with corresponding acoustic attributes--content, timbre, and prosody. 
This is achieved through the dedicated facial encoders and acoustic attribute adapters. 
Each facial encoder processes specific facial elements, and the following transposed convolution layers learn temporal alignment between visual and acoustic features.
The adapters, which include acoustic attribute predictors, align visual cues with their corresponding speech features, and adapt these features into latent sequence.
As in previous works~\cite{ren2020fastspeech2, kim2024let}, we adopt teacher-forcing strategy to train the acoustic attribute adapters.
The following paragraphs detail each modeling stage in our visual encoder.

\paragraph{Content.} 
Building on the strong correlation between lip movements and speech content~\cite{hsu2022u,shi2022learning}, we begin with content modeling by focusing on the lip motions in silent video.
We extract lip motion features through AV-HuBERT~\cite{shi2022learning}, which has been demonstrated to offer a powerful acoustic representations from lip movements~\cite{hsu2023revise,choi2023intelligible}.
Considering the fact that hidden features from each layer of the AV-HuBERT capture distinct aspects of speech~\cite{pasad2023comparative}, we integrate all these features through a learnable weighted summation.
This allows the model to learn the optimal combination of the intermediate AV-HuBERT features, enhancing the capability of the resulting representation~\cite{huang2022investigating, tsai2022superb}.

\begin{figure}[t]
    \centering
    \includegraphics[width=0.8\columnwidth]{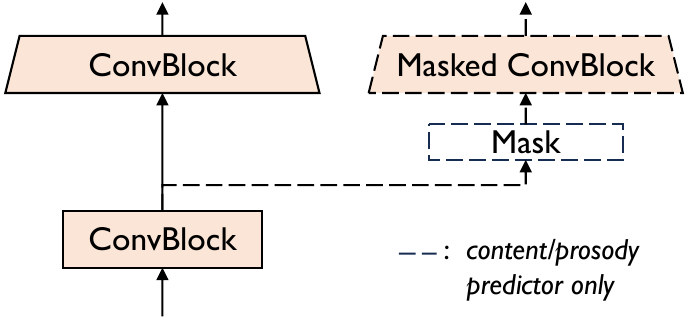}
    \caption{Speech attribute prediction pipeline. 
    The content and prosody predictor incorporate an auxiliary masked convolution block to enrich contextual information.}
    \label{fig:maskedpred}
    \vspace{-3mm}
\end{figure}

The following content adapter, which includes a content predictor, strengthens the correlation between lips and contents, while enriching content information to the hidden sequence.
The target content sequence of the predictor is obtained from the last layer of the HuBERT~\cite{hsu2021hubert} and subsequently quantized by the K-means algorithm (i.e., speech units).
In addition to convolution blocks ($CP$)~\cite{woo2023convnext}, the content predictor incorporates an auxiliary masked convolution block ($CP_m$)~\cite{liu2020non}, as illustrated in Figure~\ref{fig:maskedpred}.
This block estimates the target value at a certain frame from adjacent frames, allowing the model to learn temporal dependencies across the sequence.
We optimize the content predictor by using Cross Entropy (CE) loss with label smoothing, which is defined as:
\begin{equation}
    \begin{split}
    &\mathcal{L}_{c} = \alpha\{\text{CE}({\bf c}, {CP}({\bf h}_{l})) + \text{CE}({\bf c}, {CP}_{m}({\bf h}_{l}))\}\\
     &+(1 \text{-} \alpha)\{\text{CE}({\bf u},{CP}({\bf h}_{l}))+\text{CE}({\bf u},{CP}_{m}({\bf h}_{l}))\},
    \end{split}
\end{equation}
where ${\bf c}$, ${\bf h}_{l}$, and ${\bf u}$ denote the target content sequences, the hidden lip features, and the uniform distribution, respectively.
The label smoothing parameter $\alpha$ is set to $0.9$.

The content sequences are embedded via an learnable embedding table and then added to the hidden sequence.
These content-adapted features, which serve as the basis for the remaining hierarchical speech modeling, are passed to the subsequent timbre modeling module through the Content-to-Timbre (C2T) mapper.

\paragraph{Timbre.} 
Timbre, similar to face, is a distinct personal characteristic that specifies one's identity~\cite{joassin2011cross,mavica2013matching}.
Based on the findings that reveal the biometric relation between facial appearance and timbre~\cite{nagrani2018seeing,ning2021disentangled}, we leverage face identity to model timbre.
ArcFace~\cite{deng2019arcface} is utilized to extract discriminative face identity embeddings, which are used to predict timbre in combination with the output of C2T mapper (${\bf h}_{c2t}$).

The time-averaged feature from the first layer of HuBERT~\cite{hsu2021hubert} is used as the target timbre representation, as it is well-known for rich timbre information~\cite{fan2020exploring,choi2021neural}.
Since timbre feature does not contain temporal information, the timbre predictor relies only on a convolution pipeline ($TP$) which is optimized by Mean Absolute Error (MAE) loss:
\begin{equation}
    \mathcal{L}_{t}= \text{MAE}({\bf t}, TP({Fusion}({\bf h}_{fid},{\bf h}_{c2t}))),
\end{equation}
where ${\bf t}$ denotes the target timbre and ${\bf h}_{fid}$ represents the face identity embeddings.
The timbre value is embedded by a single linear layer and incorporated to the latent feature~\cite{lee2022pvae},
along with the output from previous level of content adapter.
The Timbre to Prosody (T2P) mapper refines this timbre-adapted feature which are then fed to the subsequent prosody modeling stage.

\paragraph{Prosody.}
Provided that prosody exhibits multiple variations even with the same contents and timbre, we regard prosody as the most complex factor which needs to be modeled at the last stage.
Specifically, we model pitch and energy sequence based on facial expression features, inspired by \cite{toisoul2021estimation,cong2023learning}.
To accurately generate prosody variations from expressions, we leverage a pre-trained facial expression encoder~\cite{zhang2024leave} that captures subtle details of expressions.

The prosody adapter comprises both pitch and energy predictors, with target values obtained from the \texttt{pYIN} algorithm~\cite{mauch2014pyin} for pitch estimation and the frequency-wise L2 norm of mel-spectrograms for energy\footnote{For robust training, the pitch value is standardized to have zero mean and unit variance over an entire sequence.}.
Similar to the content predictor, the prosody predictors consist of auxiliary masked blocks ($PP_m$) along with convolution blocks ($PP$), and are trained with their respective MAE losses:
\begin{equation}
    \begin{split}
    \mathcal{L}_{p} = & \text{MAE}({\bf p}, PP(Fusion({\bf h}_{fe}, {\bf h}_{c2p}))) \\
    & + \text{MAE}({\bf p}, PP_{m}(Fusion({\bf h}_{fe}, {\bf h}_{c2p}))),
    \end{split}
\end{equation}
where ${\bf p}$, ${\bf h}_{fe}$, and ${\bf h}_{c2p}$ refer to the target prosody sequences, the hidden features for facial expressions, and the output of T2P mapper, respectively.
Pitch and energy sequences are embedded through their respective convolution layers, and then added to hidden sequence with the timbre feature from the previous stage.

Finally, we add Transformer blocks followed by a single projection layer to yield visual encoding $\boldsymbol{\mu}$.
This encoding serves as the conditional input for the subsequent flow matching decoder, which is explained in the next section.

\subsection{Flow Matching Decoder}
We utilize a flow matching generative model as our decoder to effectively model the target mel-spectrogram distribution.
We first provide a brief overview of flow matching and then detail the architecture of our decoder.

\paragraph{Flow Matching Overview.}
Let ${\bf x}$ be a data sample from the target distribution $q({\bf x})$, and let $p_0({\bf x})$ be the simple prior distribution. 
Flow matching is a method for fitting a probability density path $p_t: [0,1] \times \mathbb{R}^d \rightarrow \mathbb{R}_{>0}$ between $p_0({\bf x})$ and $p_1({\bf x})$, which approximates $q({\bf x})$. 
Following Lipman et al.~\cite{lipman2022flow}, we define the flow $\phi_t$ as the mapping between the two distributions through the ordinary differential equation:
\begin{align}
\frac{d}{dt}\phi_t({\bf x})
& = {v}_t(\phi_t({\bf x}))
\text{,}
\qquad
\phi_0({\bf x})
= {\bf x}
\text{,}
\label{eq:ode}
\end{align}
where $t\in[0,1]$ and ${v}_t({\bf x}) = {v}_t({\bf x}; \theta)$ is the vector field parameterized by $\theta$ that specifies the trajectory of the probability flow. 
This formulation generates the probability path $p_t$, allowing us to sample from $p_t$ by solving the initial value problem.
Assume that there exist a known vector field $u_t$ that generates $p_t$. The flow matching objective aims to align $v_t({\bf x})$ with $u_t$.
In practice, however, this flow matching objective is intractable because we lack prior knowledge of $p_t$ or $v_t$. 
To address this, Lipman et al.~\cite{lipman2022flow} construct $p_t({\bf x})$ via a mixture of simpler conditional paths, for which the vector field can be easily computed.

In our case, we utilize simple optimal transport path as our conditional flow to ensure effective and efficient training~\cite{mehta2024matcha}. 
Consequently, our Optimal Transport Conditional Flow Matching (OT-CFM) loss can be defined:
\begin{equation}
\begin{split}
\mathcal{L}_{OT-CFM}(\theta) =&\mathbb{E}_{t, q({\bf x}_1), p_0({\bf x}_0)}\Vert{u}^{\text{OT}}_t(\phi^\text{OT}_t({\bf x}_0)| {\bf x}_1)\\
&-{v}_t(\phi^{\text{OT}}_t({\bf x}_0) | \boldsymbol{\mu}; \theta) \Vert^2,
\end{split}
\end{equation}
where ${\bf x}_0$ and ${\bf x}_1$ are data samples from $p_0({\bf x})$ and $p_1({\bf x})$, respectively.
The flow is defined as $\phi^{\text{OT}}_{t}({\bf x}) = (1- (1-\sigma_{\mathrm{min}})t){\bf x}_0 + t {\bf x}_1$, then the conditional target vector field is given by $\boldsymbol{u}^{\text{OT}}_t(\phi^{\text{OT}}_t({\bf x}_0)\vert {\bf x}_1) = {\bf x}_1-(1-\sigma_{\mathrm{min}}){\bf x}_0$.
Due to the linear trajectory, this achieves superior performance with fewer sampling steps compare to score-based models~\cite{ho2020denoising}.

\paragraph{Decoder Architecture.}
Our decoder is based on a U-Net architecture incorporating residual 1D convolution blocks followed by a Transformer block with snake beta activation function~\cite{lee2022bigvgan,mehta2024matcha}.
For better sampling quality, we incorporate a negative log-likelihood encoder loss~\cite{popov2021grad,mehta2024matcha}, which can be defined as follows:
\begin{equation} 
\label{eq:loss_enc}
    \mathcal{L}_{enc} = -\sum_{i=1}^{T}{\log{\varphi({\bf x}_i;\boldsymbol{\mu}_{i}, I)}},
\end{equation}
where $\varphi(\cdot;\boldsymbol{\mu}_{i},I)$ is a probability density function of $\mathcal{N}(\boldsymbol{\mu}_{i}, I)$, and $T$ denotes the temporal length.
To summarize, the total loss function $\mathcal{L}_{total}$ is defined as follows:
\begin{equation}
    \mathcal{L}_{total} = \mathcal{L}_{OT-CFM}+\mathcal{L}_{enc} + \lambda_c\mathcal{L}_{c} + \lambda_t\mathcal{L}_{t} + \lambda_p\mathcal{L}_{p},
\end{equation}
where $\lambda_c$, $\lambda_t$, and $\lambda_p$ are set to 0.5 in our experiments.

Moreover, to further enhance conditional probability path, we incorporate Classifier-Free Guidance (CFG)~\cite{ho2021classifier} which has demonstrated its effectiveness in improving generation quality~\cite{nichol2021improved,kim2024p}.
During training, we randomly drop the conditional input ($\boldsymbol{\mu}$) with a fixed probability of 0.1.
In inference, the speech decoder iteratively refines ${\bf x}_t$ with a step size of $\epsilon$, directing the trajectory away from the unconditional flow.
We employ an Euler solver with CFG: 
\begin{equation}
    \begin{split}
    {\bf x}_{t+\epsilon} = {\bf x}_t + \epsilon \{
    (1&+\beta)\cdot v_t(\phi^{\text{OT}}_{t}(\bf x)|\boldsymbol{\mu};\theta) \\
    &-\beta\cdot v_t(\phi^{\text{OT}}_{t}(\bf x)|\varnothing;\theta)
    \},
    \end{split}
\end{equation}
where $\beta$ denotes the guidance scale for CFG.
\section{Experimental Settings}
\subsection{Datasets}
\paragraph{LRS3-TED}\cite{afouras2018lrs3} is a well-established dataset for evaluating VTS systems. 
It includes approximately 440 hours of video clips sourced from TED and TEDx talks, featuring thousands of speakers and over 50,000 words.
We split the dataset in accordance with previous works~\cite{mira2022svts,choi2023intelligible,choi2023diffv2s}, ensuring no speaker overlap between the training and test sets.

\paragraph{LRS2-BBC}\cite{son2017lip} is a large-scale and real-world video dataset, which comprises 224 hours of video from BBC television shows. 
To assess the generalization capability across different datasets, we also evaluate our model on the LRS2 dataset. 
It is important to note that all models are trained exclusively on the LRS3 dataset, while the LRS2 dataset is used solely for test dataset.

\subsection{Preprocessing}
We crop face sequences from 25 fps video using RetinaFace~\cite{deng2020retinaface} and extract facial landmarks with FAN~\cite{bulat2017far}.
Lip frames are then extracted based on these landmarks and converted to grayscale.
The corresponding 16 kHz audio is transformed into a log-scale mel-spectrogram with a hop size of 320, window size of 1280, and 80 mel bins, resulting in a fixed 1:2 length ratio between the video and mel-spectrogram.
We use pre-trained HuBERT (Large)\footnote{\url{https://huggingface.co/facebook/hubert-large-ll60k}} to obtain the target content and timbre features.
Content features are then quantized by K-Means algorithm, trained on LJ Speech dataset~\cite{ljspeech17}, with 1,000 clusters.

\subsection{Implementation Details}
In our visual encoder, AV-HuBERT (Large)\footnote{\url{https://github.com/facebookresearch/av_hubert}} is used for lip encoder, and all Transformer blocks consist of 2 Transformer layers with 4 attention heads and a latent dimension of 512.
For the fusion module, we concatenate two distinct latent features along the channel dimensions and project them using 2 convolution blocks.
The configuration of our flow matching decoder follows that of Matcha-TTS~\cite{mehta2024matcha} with $\sigma_{\mathrm{min}}$ set to 10$^{-4}$.
Additionally, we adopt cosine scheduling strategy for the time-step $t$~\cite{du2024cosyvoice}.

Our model is trained on four NVIDIA A5000 GPUs with a batch size of 64.
We use AdamW optimizer~\cite{loshchilov2017decoupled} with $\beta_1$ = 0.8, $\beta_2$ = 0.99, and $\epsilon$ = $10^{-9}$.
The initial learning rate is set to $10^{-4}$, with a decay rate of $0.999^{1/8}$.
Random 96 consecutive frames are used during training, and the model is trained for 350K steps.
For robust training, we apply data augmentation to lip frames, as in previous works~\cite{mira2022svts,choi2023intelligible, kim2024let}.

\subsection{Evaluation Metrics}
The generation performance is evaluated through both subjective and objective metrics.
For subjective evaluation, we conduct 5-scale Mean Opinion Score (MOS) tests, where 25 domain-expert subjects rate the quality of 40 speech samples in terms of naturalness and intelligibility.
In the naturalness test, subjects are asked to focus on the audio quality, while in the intelligibility test, they assess the clarity of the speech content.
For objective metrics, we measure UTMOS~\cite{saeki2022utmos} and DNSMOS~\cite{reddy2021dnsmos}, which are widely used  networks to estimate perceptual audio quality~\cite{richter2023speech,siuzdak2023vocos}.
We also calculate the root mean square of F0 (RMSE$_{f0}$) to assess pitch accuracy and the Word Error Rate (WER) to evaluate the intelligibility.
WER quantifies the differences between the ground truth text labels and speech recognition results obtained from Whisper (Medium)~\cite{radford2023robust}.

\subsection{Baseline Methods}
Our method is compared to several state-of-the-art methods: SVTS~\cite{mira2022svts}, Intelligible~\cite{choi2023intelligible}, LTBS~\cite{kim2024let}, and DiffV2S~\cite{choi2023diffv2s}.
We follow the official implementations for Intelligible~\cite{choi2023intelligible}, LTBS~\cite{kim2024let}, and DiffV2S~\cite{choi2023diffv2s}, as provided by the authors.
Regarding SVTS, the LRS3 test samples are provided by the authors, while the LRS2 test samples are generated from our own reproduction, based on the official implementation of Intelligible\footnote{\url{https://github.com/choijeongsoo/lip2speech-unit}}.
Note that both SVTS and Intelligible use speaker embeddings derived from reference speech, while LTBS, DiffV2S, and our method estimate speaker characteristics directly from the silent video.

\begin{table}[t]
\centering
\resizebox{0.95\columnwidth}{!}
{
\begin{tabular}{lccc}
\toprule
Method & Naturalness$\uparrow$ & Intelligibility$\uparrow$     \\ \cmidrule(lr){1-3}
    Ground Truth &4.54 $\pm$ 0.12 &4.84 $\pm$ 0.06  \\ \cmidrule(lr){1-3}
    \multicolumn{3}{l}{\textit{Audio-driven speaker embedding}}\\
    SVTS~\cite{mira2022svts} &1.10 $\pm$ 0.06 &1.66 $\pm$ 0.14  \\ 
    Intelligible~\cite{choi2023intelligible} &2.42 $\pm$ 0.18 &3.40 $\pm$ 0.20  \\ \cmidrule(lr){1-3}
    \multicolumn{3}{l}{\textit{Video-driven speaker embedding}}\\
    LTBS~\cite{kim2024let} &2.52 $\pm$ 0.14 &2.10 $\pm$ 0.15  \\
    DiffV2S (1000)~\cite{choi2023diffv2s} &2.97 $\pm$ 0.17 &3.16 $\pm$ 0.19 \\ 
    \textbf{Ours} (10) &{\bf 4.49} $\pm$ {\bf 0.11} &{\bf 4.01} $\pm$ {\bf 0.15} \\ 
    \bottomrule
\end{tabular}
}
\vspace{-2mm}
\caption{Subjective evaluation results on LRS3 test dataset.
The results are presented with 95\% confidence interval.
The number in parenthesis means the number of sampling steps.}
\label{table:subjective}
\end{table}

\begin{table*}[ht]
\centering
\resizebox{\textwidth}{!}
{
\begin{tabular}{lccccccccc}
\toprule
\multirow{2}{*}{Method}   &\multirow{2}{*}{Steps}  &\multicolumn{4}{c}{\bfseries LRS3-TED} & \multicolumn{4}{c}{\bfseries LRS2-BBC} \\ \cmidrule(lr){3-6}\cmidrule(lr){7-10}
      & &UTMOS$\uparrow$ &DNSMOS$\uparrow$ &RMSE$_{f0}$$\downarrow$ &WER$\downarrow$  &UTMOS$\uparrow$ &DNSMOS$\uparrow$ &RMSE$_{f0}$$\downarrow$ &WER$\downarrow$ \\ \cmidrule(lr){1-10}
Ground Truth    &--  &3.545 &2.582 &-- &~~2.29 &3.013 &2.256 &-- &~~8.93  \\ \cmidrule(lr){1-10}
\multicolumn{9}{l}{\textit{Audio-driven speaker embedding}} \\
SVTS~\cite{mira2022svts}  &-- &1.283 &1.860 &56.929 &84.98 &1.387 &1.434 &53.475 &83.38 \\
Intelligible~\cite{choi2023intelligible}  &-- &2.702 &2.395 &{39.377} &{\bf 29.60} &2.331 &2.000 &{\bf 41.233} &39.53 \\ \cmidrule(lr){1-10}
\multicolumn{9}{l}{\textit{Video-driven speaker embedding}} \\
LTBS~\cite{kim2024let}  &-- &2.417 &2.361 &40.006 &84.08 &2.288 &2.174 &43.653 &94.25 \\
DiffV2S~\cite{choi2023diffv2s} &1000 &{3.058} &{2.558} &40.893 &41.07 &2.945 &2.363 &{44.414} &54.86 \\
\textbf{Ours} &10 &{\bf 4.031} &{\bf 2.789} &\underline{39.013} &{30.45} &{\bf 3.921} &{\bf 2.586} &\underline{43.441} &\underline{39.37} \\
\textbf{Ours} &1000 &\underline{3.993} &\underline{2.759} &{\bf 38.928} &\underline{30.37} &\underline{3.881} &\underline{2.552} &43.702 &{\bf 39.05} \\ \bottomrule
\end{tabular}
}
\vspace{-2mm}
\caption{Results of objective evaluation on both LRS3 and LRS2 test datasets. $\uparrow$ denotes higher is better, and $\downarrow$ means lower is better. 
{Bold} and {underlined} values represent the best and second-best results, respectively.}
\label{table:objective}
\vspace{-3mm}
\end{table*}
\section{Experimental Results}
\subsection{Subjective Evaluation}
To examine the perceptual quality of our method, we perform subject MOS tests which are regarded as the gold standard for evaluating speech generation systems~\cite{lo2019mosnet,maiti2023speechlmscore}.
MOS tests are conducted on the LRS3 test set, focusing on two key criteria: naturalness and intelligibility.
As demonstrated in Table~\ref{table:subjective}, our method produces high-quality speech, significantly outperforming existing methods on both naturalness and intelligibility.
Furthermore, our method closely approximates the naturalness of the ground truth speech with a minimal gap of only \textit{0.05}.
This indicates that the speech generated by our method is almost indistinguishable from real human recordings in terms of perceptual audio quality.

\subsection{Objective Evaluation}
In addition to subjective MOS tests, we compute UTMOS, DNSMOS, RMSE$_{f0}$, and WER, as objective evaluation metrics.
As shown in Table~\ref{table:objective}, our method shows clear improvements over standard VTS systems on both the LRS3 and LRS2 datasets, indicating that our method successfully reduces the modality gap between video and speech.
Notably, our method achieves the best audio quality across all datasets, as measured by UTMOS~\cite{saeki2022utmos} and DNSMOS~\cite{reddy2021dnsmos}, even exceeding those of ground truth audio. 
This can be attributed to the fact that our approach generates clean speech solely from face sequences, excluding background noises. 
In contrast, real-world ground truth audio often contains significant background noise, which adversely affects the audio quality.
Furthermore, the small quality difference in our method between using 10 and 1000 sampling steps confirms that the flow matching decoder can produce high-fidelity results with only a few sampling steps.

\subsection{Analysis on Speaker Similarity}
We evaluate the robustness of video-driven speaker representations to determine whether the video-driven embeddings capture accurate speaker characteristics when compared to ground truth audio.
To do this, we compute Speaker Embedding Cosine Similarity (SECS) between the speaker representations from the target and synthesized audio, using all samples from the LRS3 test set.
For an accurate and comprehensive analysis, we extract speaker representations using two different methods: GE2E~\cite{wan2018generalized}, a widely used speaker verification model for evaluating speaker similarity~\cite{choi2023diffv2s}, and VoxSim~\cite{ahn2024voxsim}, designed specifically to estimate perceptual voice similarity. 
As shown in Table~\ref{table:secs}, our method achieves the best SECS scores in cases of both GE2E and VoxSim embeddings.
This indicates that our video-driven embeddings capture precise speaker characteristics, making the voice of the generated speech more closely resemble that of original speaker, compared to existing methods that utilize video-driven embedding.

\subsection{Mel-spectrogram Visualization}
\begin{figure*}[t]
    \centering
    \includegraphics[width=.99\textwidth]{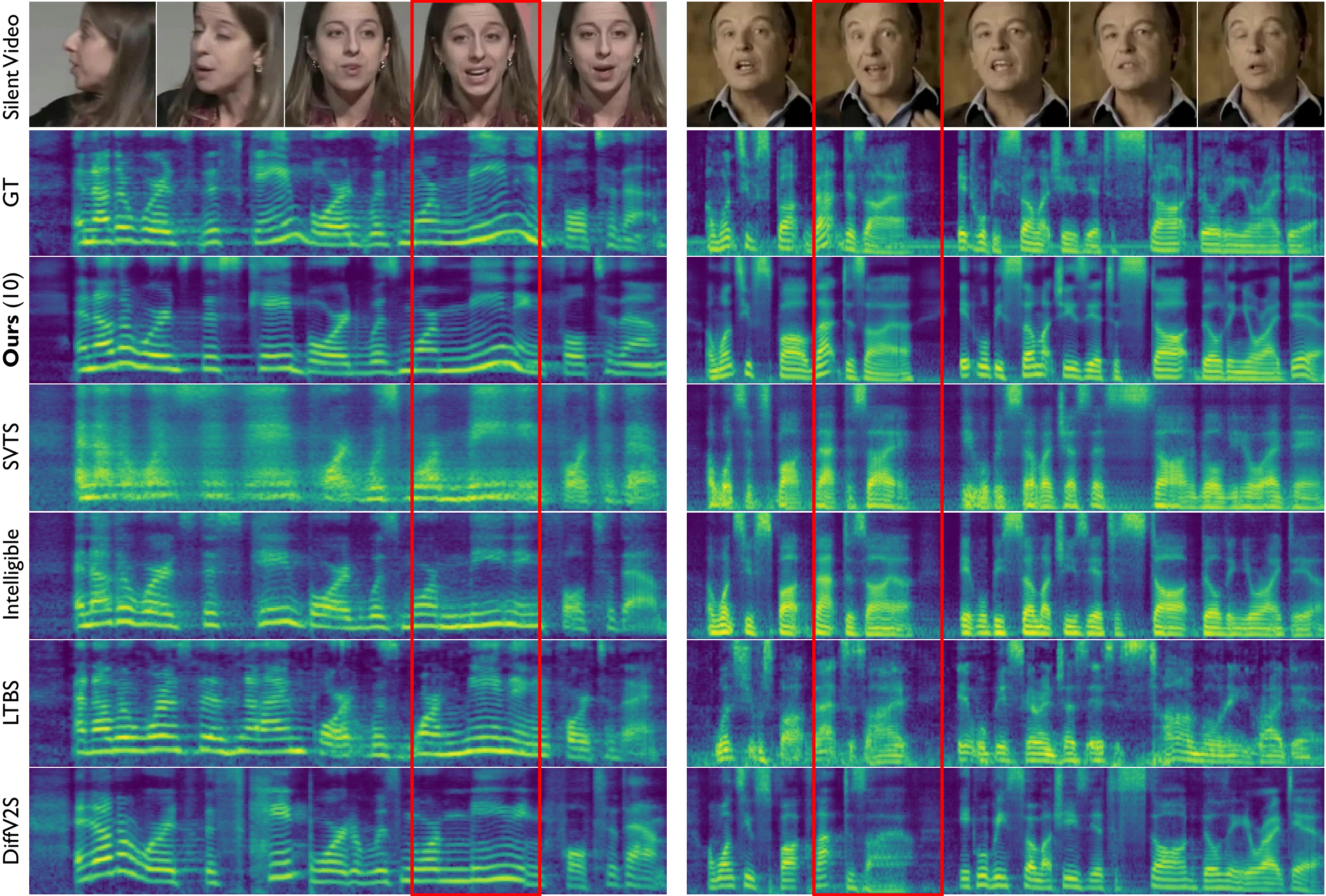}
    \vspace{-2mm}
    \caption{
    Mel-spectrogram visualization compared to Ground Truth (GT) speech. 
    As highlighted in the red boxes, the proposed method effectively captures both accurate and dynamic fundamental frequency, along with synchronized changes in facial expressions.
    }
    \label{fig:mel_plot}
\end{figure*}

For an intuitive comparison with baseline methods, we visualize the generated speech by using mel-spectrograms alongside ground truth speech.
Figure~\ref{fig:mel_plot} depicts these visualization results, where the mel-spectrogram from our system closely resembles the ground truth, capturing fine acoustic details and accurate harmonic structure. 
Additionally, we observe that our method enriches prosody by leveraging facial expressions, as reflected in the dynamic variations of the fundamental frequency along with abrupt changes in facial expressions.

\subsection{Ablation Study}
To verify the effectiveness of each component in our method, we conduct ablation studies using various metrics, including MAE$_{E}$ which refers to the MAE between the energy sequences of the target and predicted speech.
For the ablation study, we set the number of sampling steps to 10 and use the LRS3 dataset.

\paragraph{Hierarchical Modeling.}
We first explore the impact of hierarchical video-to-speech encoding, with the results presented in Table~\ref{table:ablation}.
When all mappers are removed and acoustic attributes are modeled simultaneously (\textit{w/o} Hier), the performance shows a noticeable drop across all metrics, underscoring the benefits of learning hierarchical representations between video and speech.
In our preliminary experiments, the absence of content modeling results in incomprehensible speech, implying the crucial role of content modeling in constructing a robust VTS system.
Removing the timbre (\textit{w/o} Timbre) or prosody modeling stage (\textit{w/o} Prosody), along with their respective mappers, also leads to consistent quality degradation, verifying the importance of these stages in building a high-quality VTS system.

\paragraph{Facial Features.}
The benefits of associating facial features with their acoustic counterparts are also clearly evident.
Excluding facial identity (\textit{w/o} Face ID) or expression (\textit{w/o} FE) features results in degraded quality, including declines in speaker similarity and prosody accuracy. 
This result shows the benefits of utilizing face identity and facial expressions as conditional inputs, confirming the cross-modal correlation between facial and acoustic features.

\paragraph{Training Strategy.}
We investigate the effect of weight summation across AV-HuBERT intermediate features (\textit{w/o} WS) and the masked convolution block in the content and prosody predictors (\textit{w/o} MP). 
As shown, these modules collectively contribute to improving model performance. 
In particular, integrating AV-HuBERT features through weighted summation strengthens acoustic capabilities of the representations, leading to noticeable degradation across all metrics except for a slight difference in UTMOS.

\begin{table}[t]
\centering
\resizebox{0.95\columnwidth}{!}
{
\begin{tabular}{c|cccc}
\toprule
Method &LTBS &DiffV2S &{\bf Ours} (10) &{\bf Ours} (1000)  \\ \cmidrule(lr){1-5}
 GE2E$\uparrow$  &0.609 &0.621 & {\bf 0.650} &{\bf 0.650}\\ 
 VoxSim$\uparrow$  &0.399 &0.433 &{\bf 0.495} &0.494\\ 
 \bottomrule
\end{tabular}
}
\vspace{-2mm}
\caption{SECS evaluation results on LRS3 test set. 
All speaker identities are unseen during training.}
\label{table:secs}
\end{table}

\begin{table}[t]
\centering
\resizebox{0.99\columnwidth}{!}
{
\begin{tabular}{lccccc}
\toprule
Method 
      & UTMOS$\uparrow$ &RMSE$_{f0}$$\downarrow$ &MAE$_E$$\downarrow$ &WER$\downarrow$ &GE2E$\uparrow$     \\ \cmidrule(lr){1-6}
\textbf{Ours} (10) &4.031 &39.013 &0.650 &30.45 &0.650 \\ \cmidrule(lr){1-6}
\emph{~~w/o} Hier   &3.737 &40.260 &0.763 &33.64 &0.636 \\
\emph{~~w/o} Timbre   &3.858 &39.563 &0.635 &31.15 &0.632 \\ 
\emph{~~w/o} Prosody  &3.866 &39.590 &0.677 &35.03 &0.653 \\  \cmidrule(lr){1-6}
\emph{~~w/o} Face ID &4.001 &40.751 &0.667 &31.38 &0.640 \\ 
\emph{~~w/o} FE &3.965 &40.115 &0.662 &31.08 &0.651 \\ \cmidrule(lr){1-6}
\emph{~~w/o} WS &4.038 &40.286 &0.674 &30.90 &0.649 \\  
\emph{~~w/o} MP   &3.986 &39.145 &0.650 &30.89 &0.654 \\ 
\bottomrule
\end{tabular}
}
\vspace{-2mm}
\caption{Ablation study results on the LRS3 test set. 
For brevity, we use the following abbreviations: Hier for hierarchical modeling, FE for facial expressions, WS for weighted summation in AV-HuBERT, and MP for the masked convolution prediction.}
\label{table:ablation}
\vspace{-3mm}
\end{table}

\begin{table}[t]
\centering
\resizebox{0.999\linewidth}{!}
{
\begin{tabular}{c ccccc}
\toprule
$\beta$ & UTMOS$\uparrow$ &RMSE$_{f0}$$\downarrow$  &MAE$_E$$\downarrow$ &WER$\downarrow$ &GE2E$\uparrow$ \\  \cmidrule(lr){1-6}
~~0 &3.799 &{\bf 34.797} &0.793 &25.71 &\underline{0.798}   \\
0.5 &\underline{3.944} &35.171 &0.734 &25.47 &{\bf 0.800}  \\
0.7 &{\bf 3.946} &35.290 &0.726 &{\bf 25.27} &\underline{0.798}   \\
1.0 &3.941 &\underline{34.982} &0.719 &\underline{25.38} &0.794   \\
2.0 &3.831 &35.881 &\underline{0.707} &25.78 &0.781   \\
4.0 &3.297 &37.300 &{\bf 0.674} &28.03 &0.745  \\
\bottomrule
\end{tabular}
}
\vspace{-2mm}
\caption{Analysis of guidance scale on LRS3 validation set. $\beta=0$ refers to not using classifier-free guidance.}
\vspace{-3mm}
\label{tab:cfg}
\end{table}

\paragraph{Guidance Scale.}
To find the optimal guidance scale $\beta$, we assess the model performance on the LRS3 validation set.
In Table~\ref{tab:cfg}, the benefits of applying CFG are evident in the first row ($\beta=0$; no CFG applied), where performance decreases across all metrics except RMSE$_{f0}$.
We analyze the trade-offs across various guidance scales and select $\beta = 0.7$, as it yields the best results for UTMOS and WER.

\section{Conclusion}
In this paper, we propose a novel VTS framework that generates high-quality speech from silent videos of talking faces.
We directly address the large modality gap between video and speech, and successfully mitigate the gap by learning hierarchical associations between the two modalities.
Additionally, we incorporate flow matching into the VTS system to produce realistic speech while preserving fine details.
Both subjective and objective evaluations demonstrate the superior quality of our method compared to existing approaches.
We also conduct comprehensive ablation study and validate the effectiveness of each component of our method.  

\section*{Acknowledgement}
This work was supported by the National Research Foundation of Korea (NRF) grant funded by the Korea government (MSIT) (No.RS-2023-00212845)

{
    \small
    \bibliographystyle{ieeenat_fullname}
    \bibliography{main}
}


\end{document}